\documentclass[12pt,prd,superscriptaddress,nofootinbib,preprint,showpacs,]{revtex4-1}
\usepackage{amsmath}
\usepackage{latexsym}
\usepackage{amsfonts}

\begin{document}

\title{Thermodynamics of general scalar-tensor theory with non-minimally derivative coupling}

\author{Yumei Huang}
\email{huangymei@gmail.com}
\affiliation{Institute of Physics and Electric Engineering, Mianyang Normal University, Mianyang 610021, China}
\affiliation{Department of Astronomy, Beijing Normal University, Beijing 100875, China}

\author{Yungui Gong}
\email{yggong@mail.hust.edu.cn}
\affiliation{School of Physics, Huazhong University of Science and Technology,
Wuhan 430074, China}

\begin{abstract}
With the usual definitions for the entropy and the temperature
associated with the apparent horizon, we discuss the first law of the thermodynamics on the apparent in
the general scalar-tensor theory of gravity with the kinetic term of the scalar field non-minimally coupling to Einstein tensor.
We show the equivalence between the first
law of thermodynamics on the apparent horizon and Friedmann equation for the general models,
by using a mass-like function which is equal to the Misner-Sharp mass on the apparent horizon.
The results further support the universal relationship between the first law of thermodynamics and Friedmann equation.
\end{abstract}

\preprint{1601.01035}

\maketitle

\newpage

\section{Introduction}

The area law of the entropy \cite{Bekenstein:1973ur} and the Hawking radiation \cite{Hawking:1974sw} of black holes may be clues to quantum gravity \cite{Wald:1999vt}.
The Bekenstein-Hawking entropy of black holes is equal to
one quarter of the area of the event horizon
measured in Planck units \cite{Bekenstein:1973ur,Hawking:1974sw}. In 1981, Bekenstein extended the area law of
the entropy of black holes to a weakly self-gravitating physical system in
an asymptotically flat space-time, and he proposed
the existence of a universal entropy bound \cite{Bekenstein:1980jp}.
These ideas were further generalized to the proposal of the
holographic principle \cite{'tHooft:1993gx,Susskind:1994vu,Witten:1998qj}.
The holographic principle was later realized by the AdS/CFT correspondence
which relates a gravitational
theory in $d$-dimensional anti-de Sitter space with a conformal field theory living in a
$(d-1)$-dimensional boundary space \cite{Maldacena:1997re}.
The AdS/CFT was widely applied to the study of holographic superconductors \cite{Cai:2015cya}.

Black holes are the classical solutions of Einstein equation, so the four laws of black holes \cite{Bardeen:1973gs}
may show a deep connection between gravitation and thermodynamics. For more general space-time,
one may ask whether the connection still exists. It was shown that
we can derive Einstein equation from the first law of thermodynamics
by assuming the area law of the entropy for all local acceleration horizons \cite{Jacobson:1995ab}.
In particular, for cosmological solutions with Friedmann-Robertson-Walker (FRW) metric,
we also expect the equivalence between the first law of thermodynamics
and Friedmann equations, and the relationship was indeed derived on the apparent horizon \cite{Cai:2005ra,Akbar:2006kj,Cai:2006rs,Gong:2006ma,Gong:2007md}.
For more general theories of gravity, the universal relationship between thermodynamics and gravitation was discussed extensively
\cite{Jacobson:1995ab,Hayward:1997jp,Padmanabhan:2002sha,Padmanabhan:2003gd,Cai:2005ra,Akbar:2006kj,Cai:2006rs,
Gong:2006ma,Gong:2007md,Wang:2005pk,Eling:2006aw,Kothawala:2007em,Zhou:2007pz,Gong:2006sn,
Cai:2008gw,Hayward:2008jq,Padmanabhan:2009vy,Sharif:2012zzd,Padmanabhan:2002sha,Mitra:2015nba,Helou:2015yqa,Huang:2015yva}.
For Brans-Dicke theory \cite{Brans:1961sx} and $f(R)$ gravity, a mass-like function was introduced to keep the equilibrium first
law of thermodynamics on the apparent horizon \cite{Gong:2007md}.

Brans-Dicke theory is the simplest generalization of Einstein's general relativity.
The most general scalar-tensor theory of gravity with equation of motion which contains no more than second time derivatives
in four dimensional space-time is Horndeski theory \cite{Horndeski:1974wa}. The Lagrangian of Horndeski theory is
\begin{equation}
\label{horndeskieq1}
L_H=L_2+L_3+L_4+L_5,
\end{equation}
where $L_2=K(\phi,X)$, $L_3=-G_3(\phi,X)\Box \phi$,
$$L_4=G_4(\phi,X)R+G_{4,X}\left[(\Box\phi)^2-(\nabla_\mu\nabla_\nu\phi)(\nabla^\mu\nabla^\nu\phi)\right],$$
\begin{equation}
\begin{split}
L_5=&G_5(\phi,X)G_{\mu\nu}\nabla^\mu\nabla^\nu\phi-\frac{1}{6}G_{5,X}\left[(\Box\phi)^3\right.\\
&\left.-3(\Box\phi)(\nabla_\mu\nabla_\nu\phi)(\nabla^\mu\nabla^\nu\phi)
+2(\nabla^\mu\nabla_\alpha\phi)(\nabla^\alpha\nabla_\beta\phi)(\nabla^\beta\nabla_\mu\phi)\right],
\end{split}
\end{equation}
$X=-\nabla_\mu\phi\nabla^\mu\phi/2$, $\Box\phi=\nabla_\alpha\nabla^\alpha\phi$,
the functions $K$, $G_3$, $G_4$ and $G_5$ are arbitrary functions of $\phi$ and $X$, and $G_{4,X}(\phi,X)=dG_4(\phi,X)/dX$.
The general non-minimal coupling $F(\phi)R$ is included in $L_4$ if we choose $G_4(\phi,X)=F(\phi)$.
The non-minimally derivative coupling with the kinetic term coupling to Einstein tensor is included in $L_5$
if we choose $G_5(\phi,X)=\phi$ \cite{Sushkov:2009hk,Germani:2010gm}. The scalar-tensor theory with the non-minimally
derivative coupling $\omega^2 G^{\mu\nu}\phi_{,\mu}\phi_{,\nu}$ was discussed by lots of researchers
recently \cite{Amendola:1993uh,Capozziello:1999uwa,Capozziello:1999xt,Daniel:2007kk,Germani:2011bc,Germani:2010ux,Germani:2011ua,Germani:2014hqa,Tsujikawa:2012mk,Sadjadi:2013psa,Saridakis:2010mf,Sushkov:2012za,Skugoreva:2013ooa,
DeFelice:2011uc,Sadjadi:2010bz,Sadjadi:2013uza,Minamitsuji:2013ura,Granda:2009fh,
Granda:2010hb,Granda:2011eh,deRham:2011by,Jinno:2013fka,Sami:2012uh,Anabalon:2013oea,Rinaldi:2012vy,Koutsoumbas:2013boa,
Cisterna:2014nua,Huang:2014awa,Bravo-Gaete:2013dca,Bravo-Gaete:2014haa,Bruneton:2012zk,Feng:2013pba,
Feng:2014tka,Heisenberg:2014kea,Goodarzi:2014fna,Cisterna:2015yla,Dalianis:2014sqa,
Dalianis:2014nwa,Dalianis:2015aba,Ema:2015oaa,Aoki:2015eba,Yang:2015pga,Gong:2015p,Harko:2015pma,Matsumoto:2015hua,Zhu:2015lry,Koutsoumbas:2015ekk}.

In this paper, we discuss the thermodynamics of the general scalar-tensor theory of gravity with non-minimally derivative coupling.
The paper is organized as follows. In sect. II, we
review the scalar-tensor theory with non-minimally derivative coupling and discuss the relation between the
first law of thermodynamics on the apparent horizon and Friedmann equation, and conclusions are drawn in Sect. III.

\section{The first law of thermodynamics}
In this paper, we consider the general scalar-tensor theory of gravity  with non-minimally derivative coupling
\begin{equation}
\label{action1}
S=\int d^4x\sqrt{-g}\left[\frac{F(\phi)}{16\pi G}R-\frac{1}{2}( g^{\mu\nu}-\omega^2G^{\mu\nu})\partial_{\mu}\phi\partial_{\nu}\phi-V(\phi)\right]+S_{\rm b},
\end{equation}
where $V(\phi)$ is the potential of the scalar field, the derivative coupling constant $\omega$ has the dimension of inverse mass,
$S_{\rm b}$ is the action for the matter, and the general non-minimal coupling $F(\phi)$ is an arbitrary function.
For Brans-Dicke theory, $F(\phi)=\phi$ \cite{Brans:1961sx}. For more general non-minimal coupling, we usually choose $F(\phi)=1+\xi\phi^2$,
and the special coupling $\xi=-1/6$ corresponds to the conformal coupling \cite{Callan:1970ze}.

Taking variations of the action (\ref{action1}) with respect to the metric
$g_{\mu\nu}$ leads to the field equation,
\begin{eqnarray}\label{field}
F(\phi)G_{\mu\nu}=8\pi G\left(T^{\rm b}_{\mu\nu}+T^{\rm \phi}_{\mu\nu}\right)+\nabla_{\mu}\partial_{\nu}F(\phi)-g_{\mu\nu}\hat\Box F(\phi),
\end{eqnarray}
where $\hat\Box=g^{\mu\nu}\nabla_\mu\nabla_\nu$, the energy-momentum tensor of the scalar field
\begin{equation}
\label{scaltmunu}
\begin{split}
T^\phi_{\mu\nu}=& \left[\phi_{,\mu}\phi_{,\nu}-\frac{1}{2} g_{\mu\nu}(\phi_{,\alpha})^2-g_{\mu\nu}V(\phi)\right]\\
&-\omega^2\left\{-\frac{1}{2}\phi_{,\mu}\phi_{,\nu}\,R+2\phi_{,\alpha}\nabla_{(\mu}\phi R^\alpha_{\nu)}+\phi^{,\alpha}\phi^{,\beta}R_{\mu\alpha\nu\beta}\right.\\
&+\nabla_\mu\nabla^\alpha\phi\nabla_\nu\nabla_\alpha\phi-\nabla_\mu\nabla_\nu\phi\hat\Box\phi-\frac{1}{2}(\phi_{,\alpha})^2 G_{\mu\nu}\\
&\left. +g_{\mu\nu}\left[-\frac{1}{2}\nabla^\alpha\nabla^\beta\phi\nabla_\alpha\nabla_\beta\phi+\frac{1}{2}(\hat\Box\phi)^2
-\phi_{,\alpha}\phi_{,\beta}\,R^{\alpha\beta}\right]\right\},
\end{split}
\end{equation}
$T^{\rm b}_{\mu\nu}$ is the energy-momentum tensor of the matter field,
and the total energy-momentum  tensor is
\begin{equation}\label{tf1}
T_{\mu\nu}=T^{\rm b}_{\mu\nu}+T^{\rm \phi}_{\mu\nu}.
\end{equation}

To discuss the cosmological evolution, we take the homogeneous and isotropic FRW metric
\begin{equation}
\label{FRW1}
d s^2=-dt^2+\frac{a(t)^2}{1-kr^2}dr^2+a(t)^2r^2(d\theta^2+\sin^2\theta d\varphi^2),
\end{equation}
where $k=0$, $-1$, $+1$ represents a flat, open, and closed universe respectively.
For convenience, we write the metric as the general form
\begin{equation}
\label{FRW2}
ds^2=g_{ab}dx^a dx^b+\tilde{r}^2 d\Omega^2,
\end{equation}
where $d\Omega^2=d\theta^2+\sin^2\theta d\varphi^2$ and $\tilde{r}=a(t)r$. For the FRW metric, $g_{tt}=-1$ and $g_{rr}=a^2(t)/(1-kr^2)$.
With this general metric, the $ab$ components of Einstein tensor becomes
\begin{equation}\label{Gab}
G_{ab}=-\left[2\tilde{r}\tilde{r}_{;ab}+g_{ab}(1-g^{cd}\tilde{r}_{,c}\tilde{r}_{,d}-2\tilde{r}\Box\tilde{r})\right]/\tilde{r}^{2},
\end{equation}
where the covariant derivative is with respect to the metric $g_{ab}$ and $\Box=g^{ab}\nabla_a\nabla_b$. Since
\begin{equation}\label{4F2}
\hat\Box F=\Box F+2\tilde{r}^{-1}\tilde{r}^{,a} F_{,a},
\end{equation}
eq. (\ref{field}) becomes
\begin{eqnarray}
\label{FG}
F(\phi)G_{ab}&&=-F(\phi)[2\tilde{r}\tilde{r}_{;ab}+g_{ab}(1-g^{cd}\tilde{r}_{,c}\tilde{r}_{,d}
-2\tilde{r}\Box\tilde{r})]/\tilde{r}^{2}\nonumber\\
&&=8\pi GT_{ab}+\nabla_{a}\partial_{b}F(\phi)-g_{ab} {\Box F(\phi)}-2g_{ab}\tilde{r}^{-1}\tilde{r}^{,c} F_{,c},
\end{eqnarray}
so
\begin{eqnarray}\label{FGab}
2\tilde{r}\tilde{r}_{;ab}+g_{ab}(1-g^{ab}\tilde{r}_{,a}\tilde{r}_{,b}-2\tilde{r}\Box\tilde{r})
&&=-\frac{\tilde{r}^2}{F}\left[8\pi GT_{ab}+\nabla_{a}\partial_{b}F(\phi)-g_{ab} {\Box F(\phi)}\right]\nonumber\\
&&+2 g_{ab}\tilde{r}\tilde{r}^{,c}\frac{F_{,c}}{F}.
\end{eqnarray}
After contraction, we get
\begin{eqnarray}
\label{cFGabt}
2\tilde{r}\Box\tilde{r}
=\frac{\tilde{r}^2}{F}\left[8\pi GT-{\Box F(\phi)}-\frac{4}{\tilde{r}}\,\tilde{r}^{,c} F_{,c}\right]+2(1-g^{ab}\tilde{r}_{,a}\tilde{r}_{,b}).
\end{eqnarray}
Substituting eq. (\ref{cFGabt}) into eq. (\ref{FGab}), we get
\begin{equation}
\label{cFGabt13}
2\tilde{r}\tilde{r}_{;ab}
=8\pi G\frac{\tilde{r}^2}{F}(g_{ab}T-T_{ab})-2 g_{ab}\tilde{r}\tilde{r}^{,c}\frac{F_{,c}}{F}
-\frac{\tilde{r}^2}{F}\,\nabla_{a}\partial_{b}F(\phi)+g_{ab}(1-g^{cd}\tilde{r}_{,c}\tilde{r}_{,d}).
\end{equation}

For the FRW metric (\ref{FRW1}) and a perfect fluid for the source $T_{\mu\nu}$, eq. \eqref{cFGabt13} reduces to the cosmological equations,
\begin{gather}
\label{fried1st}
3F\left(H^2+\frac{k}{a^2}\right)=8\pi G \rho-3H\dot F,\\
\label{fried2st}
2F\left(\dot H-\frac{k}{a^2}\right)=-8\pi G (\rho+p)+H\dot F-\ddot F,
\end{gather}
where the energy density $\rho$ and the pressure $p$ are
\begin{gather}
\label{rhost1}
\rho=\rho_{\rm b}+\frac{\dot\phi^2}{2}(1+{9}{\omega^2}H^2)+V(\phi),\\
\label{pst1}
p=p_{\rm b}+\frac{\dot\phi^2}{2}\left[1-{\omega^2}\left(2\dot H+3H^2+\frac{4H\ddot\phi}{\dot\phi}\right)\right]-V(\phi).
\end{gather}
Comparing eqs. \eqref{rhost1} and \eqref{pst1} with the standard Friedmann equations, we can think the extra terms as the
energy density and pressure due to the non-minimal coupling $F(\phi)$,
\begin{equation}
\label{rhopfunc1}
\rho_F=-\frac{3 H\dot F}{8\pi G},\quad p_F=\frac{\ddot F}{8\pi G}+\frac{H\dot F}{4\pi G}.
\end{equation}
With the FRW metric, the energy conservation equation becomes
\begin{equation}
\label{conflu}
8\pi G \left[\dot\rho+3H(\rho +p)\right]=3\dot F \left(2H^2+\dot H+\frac{k}{a^2}\right).
\end{equation}
In terms of the total energy $\rho_t=\rho+\rho_F$ and pressure $p_t=p+p_F$, we get
\begin{equation}
\label{conflu22}
8\pi G \left[\dot\rho_t+3H(\rho_t +p_t)\right]=3\dot F \left(H^2+\frac{k}{a^2}\right).
\end{equation}

Now let us discuss the first law of thermodynamics on the apparent horizon. The apparent horizon is
defined as $h=g^{ab}\tilde{r}_{,a}\tilde{r}_{,b}=0$, so
\begin{equation}
\label{RAH}
\tilde{r}_A=ar_A=(H^2+k/a^2)^{-1/2}.
\end{equation}
Take the time derivative of the apparent horizon $\tilde r_A$, we get
\begin{equation}
\label{dotRAH}
\dot{\tilde r}_A=-{\tilde r_A}^3H\left(\dot H-\frac{k}{a^2}\right).
\end{equation}
Take the future directed ingoing null vector field $k^a=(1,-Hr)$ which is also the (approximate) generator of the horizon,
and the Misner-Sharp mass
\begin{equation}\label{msmass}
{\cal M}=\frac{F(\phi)}{2G}\tilde{r}(1-g^{ab}\tilde{r}_{,a}\tilde{r}_{,b})=\frac{F(\phi)}{2G}\frac{\tilde{r}^3}{{\tilde r}_A^2},
\end{equation}
since $k^a\tilde{r}_{,a}=0$, we get
\begin{equation}
\label{decalm1}
k^a{\cal M}_{,a}={\cal M}_{,t}-Hr{\cal M}_{,r}=\frac{F_{,t}}{2G}\frac{\tilde{r}^3}{\tilde{r}_A^2}
+\frac{F}{G}\tilde{r}^3(-\frac{\dot{\tilde{r}}_A}{\tilde{r}_A^3}).
\end{equation}
Therefore, on the apparent horizon, we find
\begin{equation}
\label{decalm2}
-d{\cal E}=-k^c{\cal M}_{,c}dt
=-\frac{dF}{2G}\tilde{r}_A
+\frac{F(\phi)}{G}d\tilde{r}_A
\end{equation}
On the other hand, if we choose the horizon temperature $T_A=1/(2\pi \tilde{r}_A)$ and the horizon entropy
$S_A=\pi \tilde{r}_A^2F(\phi)/G$, then we get
\begin{eqnarray}\label{tdsa}
T_AdS_A&&=\frac{1}{2\pi G \tilde{r}_A }[2\pi F(\phi)\tilde{r}_A d\tilde{r}_A+\pi \tilde{r}_A^2 d F]\nonumber\\
&&=\frac{F(\phi)}{G}d \tilde{r}_A+\frac{d F}{2G}\tilde{r}_A \neq-d{\cal E}
\end{eqnarray}

As expected, with the usual definitions of the temperature $T_A=1/(2\pi \tilde{r}_A)$, the entropy $S_A=\pi \tilde{r}_A^2F(\phi)/G$
associated with the apparent horizon,
and the Misner-Sharp mass, the first law of thermodynamics on the apparent horizon does not hold
for general scalar-tensor theories of gravity \cite{Cai:2006rs}.
Without the non-minimal coupling $F(\phi)$, the first law of thermodynamics on the apparent horizon holds for the non-minimally derivative coupling \cite{Huang:2015yva}.
To overcome the problem arising from the general non-minimal coupling $F(\phi)R$, a mass-like function which is equal to the Misner-Sharp mass on the apparent horizon
was introduced \cite{Gong:2007md}. Therefore, following ref. \cite{Gong:2007md}, we use the mass-like function instead.
The mass-like function $M$ is defined as
\begin{equation}
\label{mslikest}
{M}=\frac{F(\phi)}{2G}\tilde{r}(1+g^{ab}\tilde{r}_{,a}\tilde{r}_{,b})=\frac{F(\phi)}{G}\tilde{r}-{\cal M}.
\end{equation}
On the apparent horizon, since $h=g^{ab}\tilde{r}_{,a}\tilde{r}_{,b}=0$ and $k^ah_{,a}=2\dot{\tilde{r}}_A/\tilde{r}_A^{-1}$,
so $M={\cal M}=F(\phi)\tilde{r}_A/(2G)=4\pi\tilde{r}_A^3\rho/3$.
By using the result \eqref{decalm2}, on the apparent horizon, we get
\begin{eqnarray}
\label{dem2tst}
d{E}=k^c{M}_{,c}dt&=&\frac{1}{G}k^c F_{,c}\tilde{r}_A dt-\frac{d F}{2G}\tilde{r}_A
+\frac{F}{G}d \tilde{r}_A\nonumber\\
&=&\frac{d F}{2G}\tilde{r}_A
+\frac{F}{G}d \tilde{r}_A=T_A dS_A
\end{eqnarray}
With the help of the mass-like function $M$, we show that the first law of the thermodynamics on the apparent horizon holds for the general
scalar-tensor theory of gravity with non-minimally derivative coupling. Note that all the variables in the above
equations are geometric quantities, the above relation is just a geometric identity and it is always true in FRW metric.
Next we need to show that the geometric quantities are related with physical variables and the identity
is equivalent to the Friedmann equation. Due to the extra term in the right hand side of eq. \eqref{conflu22}, we expect the energy flow rate
through the apparent horizon to be
\begin{equation}
\label{energyflow12}
\begin{split}
\frac{dE}{dt}&=\frac{4\pi \tilde{r}_A^3}{3}\frac{3\dot F}{8\pi G\tilde{r}_A^2}+4\pi\tilde{r}_A^3 H(\rho_t+p_t)\\
&=\frac{\tilde{r}_A }{2G}\dot F+\frac{\tilde{r}_A^3H}{2G}\left[8\pi G (\rho+p)-H\dot F+\ddot F\right].
\end{split}
\end{equation}
For Einstein gravity with $F(\phi)=1$, the above result recovers the standard relation $dE=4\pi\tilde{r}_A^3 H(\rho+p)dt$.

From the definition of the mass-like function \eqref{mslikest}, we get
\begin{equation}
\label{mctotst}
{M}_{,c}=\frac{(F\tilde{r})_{,c}}{G}-\frac{(F\tilde{r})_{,c}}{2G}(1-g^{ab}\tilde{r}_{,a}\tilde{r}_{,b})
+\frac{F \tilde{r}}{G}(\tilde{r}^{,a}\tilde{r}_{;ac}).
\end{equation}
Substituting the field equations (\ref{cFGabt}) and (\ref{cFGabt13}) into eq. (\ref{mctotst}), we get
\begin{eqnarray}\label{mctot123st}
{M}_{,c}&=&\frac{(F\tilde{r})_{,c}}{G}+2\pi \tilde{r}^3\frac{F_{,c}}{F}T-4\pi \tilde{r}^2(T^a_c-\delta^a_cT)\tilde{r}_{,a}-\frac{F_{,c}}{FG}\tilde{r}^2\tilde{r}^{,a} F_{,a}-\frac{F_{,a}}{G}\tilde{r}\tilde{r}^{,a}\tilde{r}_{,c}\nonumber\\
&&-\frac{F_{,c}}{2G}\tilde{r}^2\Box\tilde{r}-\frac{F_{,c}}{4FG}\tilde{r}^3\Box F
-\frac{F_{;ac}}{2G}\tilde{r}^2\tilde{r}^{,a}.
\end{eqnarray}
By using the field equation \eqref{mctot123st}, we get the energy flow through the apparent horizon
\begin{equation}
\label{dem2st}
d{E}=k^c{M}_{,c}dt=\frac{d F}{2G}\tilde{r}_A+\frac{\tilde{r}_A^3H}{2G}\left[8\pi G (\rho+p)-H\dot F+\ddot F\right]dt.
\end{equation}
The is exactly what we speculate in eq. \eqref{energyflow12}.
Combining eqs. (\ref{fried2st}) and (\ref{dotRAH}), we get
\begin{equation}
\label{radereq1}
\dot{\tilde{r}}_A=\frac{\tilde{r}_A^3H}{2F}\left[8\pi G(\rho+p)-H\dot{F}+\ddot{F}\right].
\end{equation}
Substituting eq. \eqref{radereq1} into eq. \eqref{dem2st}, we obtain
\begin{equation}
\label{dem2c1st}
dE=\frac{d F}{2G}\tilde{r}_A
+\frac{F}{G}d{\tilde r}_A=T_A dS_A.
\end{equation}
Therefore, the first law of thermodynamics on the apparent horizon is derived from Friedmann equation.
To derive Friedmann equation from the first law of thermodynamics on the apparent horizon \eqref{dem2c1st},
we combine eqs. \eqref{energyflow12} and \eqref{dem2c1st}, then we get
\begin{equation}
\label{fltofr1}
F\dot{\tilde{r}}_A=\frac{\tilde{r}_A^3H}{2}\left[8\pi G (\rho+p)-H\dot F+\ddot F\right].
\end{equation}
Substitute eq. (\ref{dotRAH}) into the above equation, we obtain eq. (\ref{fried2st}) from the first law of thermodynamics on the apparent horizon.
Combining the energy conservation equation (\ref{conflu}) and eq. (\ref{fried2st}), after integration,
we finally obtain the Friedmann equation (\ref{fried1st}).
With help of the mass-like function \eqref{mslikest} and the proposed energy flow through the apparent horizon \eqref{energyflow12},
we show the equivalence between the first law of thermodynamics on the apparent horizon and Friedmann equation.

%\begin{equation*}
%2F(\dot H-\frac{k}{a^2})=-8\pi G (\rho+p)+H\dot F-\ddot F.
%\end{equation*}

%\begin{equation*}
%3F(H^2+\frac{k}{a^2})=8\pi G \rho-3H\dot F.
%\end{equation*}

\section{Conclusions}

With the help of the mass-like function which is equal to the Misner-Sharp mass on the apparent horizon proposed in ref. \cite{Gong:2007md}, we show that the first law of thermodynamics
on the apparent horizon is a geometric identity for the general scalar-tensor theory of gravity with non-minimally derivative coupling. To discuss
the equivalence between the first law of thermodynamics on the apparent horizon and Friedmann equation, we need to connect the mass-like function
with the physical energy flow through the apparent horizon. In standard cosmology, the energy is conserved, i.e., $\dot\rho_b+3H(\rho_b+p_b)=0$, and the
energy flow through the apparent horizon is $4\pi\tilde{r}_A^3 H(\rho_b+p_b)$. In general scalar-tensor theories of gravity, we also expect that
the energy flow through the apparent horizon has the term $4\pi\tilde{r}_A^3 H(\rho_t+p_t)$. However,
in the general scalar-tensor theory of gravity with non-minimally derivative coupling, the total energy is not conserved, we have extra contribution
coming from the non-minimal coupling $F(\phi)$,
so we propose that the energy flow through the apparent horizon is given by eq. \eqref{energyflow12}. By using Friedmann equations and the definition
of the mass-like function, we show that eq. \eqref{energyflow12} gives the energy flow
through the apparent horizon. Then,
we show the equivalence between Friedmann equation and the first law of thermodynamics on the apparent horizon.
Therefore, our results further support the universal relationship between the first law of thermodynamics and Friedmann equations.

\begin{acknowledgments}
This work was supported by the National Natural Science Foundation of China (Grant Nos. 11175270 and 11475065)
and the Program for New Century Excellent Talents in University (Grant No. NCET-12-0205).
\end{acknowledgments}

%\bibliographystyle{scichina}
%\bibliography{../../book/cosmologyref}

\end{document}